\documentclass[12 pt]{article}
\usepackage{times,bm}
\usepackage[affil-it,]{authblk}
\usepackage[pagebackref=false,colorlinks=true,linkcolor=blue,citecolor=blue,urlcolor=blue]{hyperref}
\usepackage{geometry}
\usepackage{graphicx}
\usepackage{cite}
\usepackage{amsmath}
\usepackage{amsfonts}
\usepackage{amssymb}
\usepackage{color}
\usepackage{float}
\usepackage{multirow,multicol}
\usepackage{tabulary}
\usepackage[flushleft]{threeparttable}
\usepackage{pgfplots,subfigure}
\usepackage{xcolor}
\numberwithin{equation}{section}
\usepackage{tikz}
\usepackage{ulem}
\usepackage{hyperref}
\usepackage{nameref}

\usepgflibrary{arrows}
\usetikzlibrary{shapes.callouts}
\tikzset{
	level/.style   = { thick, },
	connect/.style = { dotted, red   },
	notice/.style  = { draw, rectangle callout, callout relative pointer={#1} },
	label/.style   = { text width=2cm }
}

\usepackage{letltxmacro}
\makeatletter
\let\oldr@@t\r@@t
\def\r@@t#1#2{%
	\setbox0=\hbox{$\oldr@@t#1{#2\,}$}\dimen0=\ht0
	\advance\dimen0-0.2\ht0
	\setbox2=\hbox{\vrule height\ht0 depth -\dimen0}%
	{\box0\lower0.4pt\box2}}
\LetLtxMacro{\oldsqrt}{\sqrt}
\renewcommand*{\sqrt}[2][\ ]{\oldsqrt[#1]{#2}}
\makeatother

\newcommand{{\ri}}{{\rm{i}}}
\newcommand{{\Psibar}}{{\bar{\Psi}}}

\begin{document}

\title{Duffin-Kemmer-Petiau particles in the presence of the spiral dislocation}
\author{\large \textit {S. Hassanabadi}$^{\ 1}$\footnote{E-mail: s.hassanabadi@yahoo.com (corresponding author)}~ ,~ \textit {S. Zare}$^{\ 2}$\footnote{E-mail: soroushzrg@gmail.com}~ ,~\textit {B. C. L\" utf\"uo\u{g}lu}$^{\ 1\,\, \& \ 3}$\footnote{E-mail: bekircanlutfuoglu@gmail.com}~,~\textit {J. K\v{r}\'i\v{z}}$^{\ 1}$\footnote{E-mail: jan.kriz@uhk.cz}~ \\and~\textit {H. Hassanabadi}$^{\ 1\,\, \& \ 2}$\footnote{E-mail: h.hasanabadi@shahroodut.ac.ir}\\
	
	\small \textit {$^{\ 1}$Department of Physics, University of Hradec Kr\'alov\'e,Rokitansk\'eho 62, 500 03 Hradec Kr\'alov\'e, Czechia}\\
	\small \textit {$^{\ 2}$Faculty of Physics, Shahrood University of Technology, Shahrood, Iran}\\
	\small \textit{P.O. Box 3619995161-316}\\
	\small \textit {$^{\ 2}$Department of Physics, Akdeniz University, Campus 07058, Antalya, Turkey,}\\
}

\date{}
\maketitle

\begin{abstract}		
In this study, we investigated the influence of the topological defects space-time with a spiral dislocation on a spin-zero boson field by using the Duffin-Kemmer-Petiau (DKP) equation. To be more specific, we solved the generalized spin-zero DKP equation in the presence of a spiral dislocation exactly. We derived the wave function and corresponding energy eigenvalues for two cases, in the absence and presence of a static potential by using analytical methods. We numerically demonstrated the effect of the spiral dislocation on the solutions.
\end{abstract}

\begin{small}
	
\end{small}

\bigskip

\newpage


\section{Introduction}
Domain wall \cite{Vilenkin1985}, global monopole \cite{BarriolaPRL1989}, and cosmic strings \cite{Vilenkin19852,cs1, cs2} as instances of topological defects in the space-time are anticipated in some unified theories of particle interactions \cite{AssisPRD2000, Furtado1999,FurtadoPRD1999}. These kinds of topological defects may have occurred at phase transitions during the universe evolution in the very early Universe \cite{Kibble1976}. A cosmic string space-time, as the best known of them, creates a bridge between the physics of microscopic and macroscopic scales. Furthermore, the cosmic strings are the linear topological defects, similar to vortex strings in superfluids or superconductors\cite{Brandenberger1992}, analogous to dislocations and disclinations in solids and condensed matter physics \cite{Katanaev}. In fact, with a clear approach, we can say, the cosmic strings are not only interested in cosmology and field theory but also interested in solids in the context of the linear topological defects. These linear topological defects, in a solid, can be characterized as distortions, which are classified as twist and wedge disclinations, and spiral and screw dislocations \cite{ZareEPJP, ahmed1, ahmed2}. It is worth mentioning that in this paper  we focus only  on the spiral dislocation.

In a series of published articles and thesis, Duffin, Kemmer, and Petiau coined a first order differential equation, hereafter (DKP),  that has similarities with the Dirac equation to deal with the meson dynamics
\cite{Kemmer1938,Duffin1938,Kemmer1939,Petiau1936}. The algebra of the DKP equation is different from the Dirac equation algebra. As a result, the commutation rules provided by the matrices of the DKP equation are also more complex. At certain time periods, interest in the DKP equation showed decreases and increases. The main reason for the decrease was that there were doubts that claim a similarity to the Klein-Gordon and Proca equations \cite {Fainberg, Bou}.  Nowadays, we know that the similarity is valid only in the very special cases \cite{Krajcik}. Therefore, in recent decades, the DKP field have been widely  studied by many researchers in Refs. \cite{Kalbermann1938, Clark1985, Daicic1992, Beckers1995, Casana2002, Lunardi2002,  Hou1, Hou2, Hou3, ZareGRG2020, ahmed3, ahmed4}. In this contribution, we analyse the influence of the topological defects space-time with a spiral dislocation on a spin-zero boson field in the absence and in the presence of a static potential by using the DKP equation. As can be seen in Refs. \cite{AssisPRD2000,FurtadoPRD1999,Furtado1999,ZareEPJP,ahmed1,ahmed2,Hou1,ZareGRG2020,ahmed3,ahmed4,FurtadoJPAMG2000,AhmedAHEP2020-2,BuenoPBCD2016,MarquesJPAMG2001,SilvaNetto,MaiaEPJC2019,ZareEPJP2020-2,Maia2019}, quantum dynamics has been studied with the motivation of investigating the topological defects theory that is defect structures in the context of Riemann–Cartan manifolds and distortions into solids in the non-relativistic and relativistic framework according to the Schr\"odinger, Dirac, DKP, and Klein–Gordon equations.
Accordingly,  Our clear motivation is finding and analyzing the energy eigenvalues and corresponding wave function of a boson particle with spin-zero in the presence of the spiral dislocation in space-time. 

This contribution is structured as follows: In section \ref{sec2}, at first, we introduce the metric of the topological defect space-time, then, we analyse the generalized spin-zero DKP equation under the background of a spiral dislocation and obtain the non-coupled second-order a differential equation by using appropriate changing variables. In section \ref{sec3}, we solve the generalized DKP equation in the absence and in the presence of a static potential energy, respectively. Afterwards, we obtain energy eigenvalues and corresponding wave functions in terms of the Bessel equation by employing the Nikiforov-Uvarov (NU) method. Finally, in section \ref{Conc}, we give a conclusion of this manuscript.




\section{Quantum dynamics in the presence of a spiral dislocation \label{sec2}}
In light of the great literature on quantum dynamics in the non-relativistic and relativistic framework in space-time \cite{AssisPRD2000,FurtadoPRD1999,Furtado1999,ZareEPJP,ahmed1,ahmed2,Hou1,ZareGRG2020,ahmed3,ahmed4,FurtadoJPAMG2000,AhmedAHEP2020-2,BuenoPBCD2016,MarquesJPAMG2001,SilvaNetto,MaiaEPJC2019,ZareEPJP2020-2,Maia2019} with dislocation fields that are considered as deformation fields associated with the dislocated space-time, we are eager to study on quantum dynamics of spin-zero bosons in the spiral dislocation space-time in the presence and  in the absence of a static potential.
It worth mentioning that the presence of deformation in space-time may cause a disconnection, which is known as dislocation, in space-time so that it leads to non-Euclidean geometry in space-time.
The non-Euclidean geometry may be demonstrated through a non-Euclidean metric.  Let us now represent space-time associated with some sort of linear topological defects (the spiral dislocation) but before that, we want to provide a classification of the linear topological defects. In this way,  the linear topological defects can be categorized into two kinds in the elastic medium background: one of them is dislocations ( spiral and screw dislocations) and the other one is disclinations (twist and wedge disclinations); meanwhile, both of which are analogous to the topological defects in the universe.
In fact, spiral dislocations (edge dislocation) are a kind of linear topological defect in which distortion of a circle into a spiral is seen. If we assume the units $\hbar=1$ and $c=1$,  then, we can express the linear topological defect which is the background of this work with the line element as follows: \cite{MaiaEPJC2019,ZareEPJP2020-2}
\begin{equation}\label{LE1}
ds^2=-dt^2+dr^2+2\beta dr d\varphi+\left(\beta^2+r^2\right)d\varphi^2+dz^2.
\end{equation}
Here, $r$, and $\varphi$ are the radial and azimuthal coordinates, respectively. The ranges of the line coordinate elements can be written as $0<t<\infty$, $0<r$, so that $r=\sqrt{x^2+y^2}$, and also $0<\varphi<2\pi$ and $-\infty<z<\infty$. The constant parameter associated with the distortion of the topological defects space-time is indicated by $\beta$. As shown in Ref. \cite{ZareEPJP2020-2,Maia2019}, the $\beta$ parameter corresponds to the Burger vector $\vec{b}$, that is, $\beta={|\vec{b}|}/2\pi$. It is worth mentioning that the spiral dislocation has the Burger vector which is parallel to the plane $z=0$. Meanwhile, the ranges of the Burger vector is $0<\beta<1$ that it is of a order of the inter-atomic displacement in a solid. According to Eq. \eqref{LE1}, the metric tensor $g_{\mu \nu}$ and the inverse metric tensor $g^{\mu \nu}$ can be written, as follows:
\begin{equation}\label{mettens}
g_{\mu\nu}=\begin{pmatrix}
-1&0&0&0\\
0&1&\beta&0\\
0&\beta&\beta^2+r^2&0\\
0&0&0&1
\end{pmatrix}, \qquad g^{\mu\nu}=\begin{pmatrix}
-1&0&0&0\\
0&1+\frac{\beta^2}{r^2}&-\frac{\beta}{r^2}&0\\
0&-\frac{\beta}{r^2}&\frac{1}{r^2}&0\\
0&0&0&1
\end{pmatrix}.
\end{equation}
The indices in the metric tensor, $\mu$ and $\nu$, are used to indicate the space-time coordinates, hence,  $\mu,\nu=t,r,\varphi,z$. Since we deal with a background of a space-time with a spacelike dislocation (spiral dislocation), we shall initially constitute a local reference frame. Considering  the metric given in Eq. \eqref{LE1}, and recalling
 $\hat{\theta}^{a}=e^{a}_{\mu}\left(x\right)dx^{\mu}$, where $a=0,1,2,3$, we obtain
\begin{equation}\label{noncor}
\begin{split}
&\hat{\theta}^0=\mathrm{d}t, \qquad \quad
\hat{\theta}^1=\mathrm{d}r+\beta \mathrm{d}\varphi,\\
&\hat{\theta}^2=r\mathrm{d}\varphi, \qquad \,\,
\hat{\theta}^3=\mathrm{d}z.\\
\end{split}
\end{equation}
Thereby, through the local reference frame, we find the components of the non-coordinate basis and their inverses, which are called as tetrads and inverse tetrads, respectively.
\begin{equation}\label{tetrads}
e_{\mu}^{a}=\begin{pmatrix}
1&0&0&0\\
0&1&\beta&0\\
0&0&r&0\\
0&0&0&1
\end{pmatrix}, \qquad e^{\mu}_{a}=\begin{pmatrix}
1&0&0&0\\
0&1&-\frac{\beta}{r}&0\\
0&0&\frac{1}{r}&0\\
0&0&0&1
\end{pmatrix}.
\end{equation}
Note that the tetrads that are obtained out of the local reference frame have to satisfy the following conditions:
\begin{equation}\label{cond1}
e_{a}^{\mu}\left(x\right)e_{\nu}^{a}\left(x\right)=\delta_{\nu}^{\mu}, \qquad
e_{\mu}^{a}\left(x\right)e_{b}^{\mu}\left(x\right)=\delta_{b}^{a},
\end{equation}
and
\begin{equation}
g_{\mu \nu}\left(x\right)=e_{\mu}^{a}\left(x\right)e_{\nu}^{b}\left(x\right)\eta_{a b},
\end{equation}
where the symbol $\eta_{a b}=\mathrm{diag}(-,+,+,+)$  represents the Minkowski space-time metric tensor.

To study the quantum dynamics of a relativistic spin-zero particle in a defected space-time background, we need to transform the partial derivative into the covariant derivative in the DKP equation. Therefore, we express the covariant derivative in the form of $\partial_{\mu}\rightarrow\nabla_{\mu}$, and then, we employ $\nabla_{\mu}=\partial_{\mu}-\Gamma_{\mu}(x)$, where  $\Gamma_{\mu}(x)$ is the affine connection. It is worth noting that in the DKP equation, one can deduce the affine connection terms out of
\begin{equation}\label{affconc}
\Gamma_{\mu}(x)=\frac12\omega_{\mu a b}\left[\beta^a,\beta^b\right], \qquad a,b=0,1,2,3.
\end{equation}
Here,  $\beta^a$ are the DKP matrices in the Minkowski space-time. As pointed out in Refs. \cite{Nedjadi1994,Hpour2018,ZareGRG2020}, by using the $\beta^a$ matrices, one can obtain the DKP algebra which has three irreducible representations. Indeed,  the ten and five dimensional representations are related to spin-one and spin-zero particles, respectively. The one-dimensional representation is the trivial one. In this contribution we examine the dynamics of the spin-zero particles, therefore we only utilize the five-dimensional representation. We choose the following $5\times5$ DKP matrices:
\begin{equation}\label{8}
\beta^0=\begin{pmatrix}
\theta&0_{2\times 3}\\
0_{3\times 2}&0_{3\times 3}
\end{pmatrix},\qquad
{\vec{\beta}}=\begin{pmatrix}
0_{2\times 2}&\vec{\tau}\\
-\vec{\tau}^T&0_{3\times 3}
\end{pmatrix},
\end{equation}
where $T$ denotes the matrix transposition, and
\begin{equation}\label{9}
\theta=\begin{pmatrix}
0&1\\
1&0
\end{pmatrix},\quad \tau^1=\begin{pmatrix}
-1&0&0\\
0&0&0
\end{pmatrix},\quad \tau^2=\begin{pmatrix}
0&-1&0\\
0&0&0
\end{pmatrix}, \quad \tau^3=\begin{pmatrix}
0&0&-1\\
0&0&0
\end{pmatrix}.
\end{equation}
Before we proceed to the calculations, it would be appropriate to state that the DKP matrices satisfy the following DKP algebra.
\begin{equation}\label{DKPAlg}
\beta^{a}\beta^{b}\beta^{c}+\beta^{c}\beta^{b}\beta^{a}=\eta^{ab}\beta^{c}+\eta^{bc}\beta^{a}, \qquad a,b,c=0,1,2,3.
\end{equation}
To get the non-null component of the affine connection, according to Eq. \eqref{affconc}, at first we need to obtain the non-null components of spin connection in the absence of torsion by solving the Maurer-Cartan structure equations, $\mathrm{d}\hat{\theta}^a+\omega^{a}_{b}\wedge\hat{\theta}^{b}=0$, in which $\omega^{a}_{b}=\omega^{a}_{\mu\,\,b}\mathrm{d}x^{\mu}$. After a simple algebra  we find $\omega^{2}_{\varphi\,\,1}=-\omega^{1}_{\varphi\,\,2}=1$. Thus, we express the affine connection as
\begin{equation}\label{affconn}
\Gamma_{\varphi}=
\begin{pmatrix}
0&0&0&0&0\\
0&0&0&0&0\\
0&0&0&1&0\\
0&0&-1&0&0\\
0&0&0&0&0
\end{pmatrix}.
\end{equation}
Then, we consider the minimal coupling by $\partial_{\mu}\rightarrow \partial_{\mu}+iqA_{\mu}$, and write the covariant derivative belonging to the generalized DKP equation in the space-time with topological defects as
\begin{equation}
\nabla_{\mu}\rightarrow \partial_{\mu}+iqA_{\mu}-\Gamma_{\varphi}.
\end{equation}
Here, $q$ is the electric charge and $A_{\mu}=(A_t,\vec{A})$ is the electromagnetic 4-vector potential, while its spatial components are $\vec{A}=(A_r,A_\varphi,A_z)$.
Thus, the generalized DKP equation for a free spin-zero boson of the mass $M$ can be written as
\begin{equation}\label{DKPEq}
\left[i\beta^{\mu}\partial_{\mu}-i\beta^{\varphi}\Gamma_{\varphi}-q\beta^{\mu}A_{\mu}-M\right]\Psi\left(t,\vec{r}\right)=0,
\end{equation}
in which the DKP field is denoted by $\Psi\left(t,\vec{r}\right)$ and the DKP matrices in the defected space-time are represented by $\beta^{\mu}$. After all, in the following equation, we show how beta matrices are related to their Minkowski space-time counterparts
\begin{equation}\label{defbeta}
\beta^{\mu}=e^{\mu}_{a}\beta^{a}.
\end{equation}
We recall the specific tetrads bases which are presented in Eq. \eqref{tetrads}, and employ them in Eq. \eqref{defbeta}. We find the defected space-time DKP-beta-matrices, $\beta^{\mu}$, in terms of the flat space-time DKP-beta-matrices, $\beta^{a}$, as follows:
\begin{equation}\label{def-beta}
\beta^{t}=\beta^{0}, \qquad \beta^{r}=\beta^{1}-\frac{\beta}{r}\beta^{2}, \qquad \beta^{\varphi}=\frac{\beta^{2}}{r}, \qquad
\beta^{z}=\beta^{3}.
\end{equation}
Let us now present the four-current associated with the DKP equation as
\begin{equation}
J^{\mu}=\frac12\bar{\Psi}\beta^{\mu}\Psi.
\end{equation}
Here, the adjoint DKP spinor $\bar{\Psi}$ is defined as $\bar{\Psi}={\Psi}^{\dagger}\eta^0$, with $\eta^0=2\beta^0\beta^0-I$ and $\Psi^{\dagger}=(\Psi^{\ast})^\dagger$ (please note that the Hermitian conjugate of $\Psi$ is denoted by $\Psi^{\dagger}$).  $I$ is the $5\times5$ identity matrix and $\beta^{0}$ is the
DKP matrix in Minkowski space-time.  Because of being Hermitian the matrices $\beta^\mu$ with regard to  $\eta^0$, we can write $\left(\eta^0 \beta^\mu\right)^{\dagger}=\eta^0 \beta^\mu$. Now, by considering the time component of $J^{\mu}$, we find the charge density, $J^{0}$, as
\begin{equation}\label{J0}
J^{0}=\frac12\Psi^{\dagger}\eta^0\beta^0\Psi.
\end{equation}
For the continuation of the process, we determine the 4-vector potential components. Henceforth, $A_{\mu}=(A_t,0)$. Then, we utilize Eqs. \eqref{affconn} and \eqref{def-beta} to obtain the second term of the left-hand side of Eq. \eqref{DKPEq} as
\begin{equation}\label{betaGamm}
-i\beta^{\varphi}\Gamma_{\varphi}=\begin{pmatrix}
0&0&-\frac{i}{r}&0&0\\
0&0&0&0&0\\
0&0&0&0&0\\
0&0&0&0&0\\
0&0&0&0&0
\end{pmatrix}.
\end{equation}
To solve the generalized DKP equation, given in Eq. \eqref{DKPEq}, we employ the spin-zero DKP spinors in terms of the eigenvalues of the z-component of the total angular momentum operator $\hat{J}_{z}=-i\partial_{\varphi}$ and the z-component of the momentum operator $\hat{p}_{z}=-i\partial_{z}$ as
\begin{equation}\label{DKPSpinor}
\Psi\left(t,\bar{r}\right)=e^{-i\mathcal{E}t+im\phi+ikz}\begin{pmatrix}
\psi_{1}\left( r\right)\\
\psi_{2}\left( r\right)\\
\psi_{3}\left( r\right)\\
\psi_{4}\left( r\right)\\
\psi_{5}\left( r\right)
\end{pmatrix}.
\end{equation}
Here, the energy of the spin-zero boson is given by $\mathcal{E}$, while the angular momentum quantum number is denoted by $m=0,\pm1,\pm2,\dots$, and the momentum along the $z$-axis is indicated by $k$. We continue the processes by substituting Eqs. \eqref{def-beta}, \eqref{betaGamm} and \eqref{DKPSpinor} within Eq. \eqref{DKPEq}, so we find the following five equations, each one associated with a component of the DKP spinor:
\begin{subequations}
\begin{eqnarray}
-M\psi_{1}(r)+\left(-qA_{t}+\mathcal{E}\right)\psi_{2}(r)-i\partial_{r}\psi_{3}(r)-\frac{i}{r}\psi_{3}(r)\nonumber\\
+\frac{m}{r}\psi_{4}(r)+\frac{i\beta}{r}\partial_{r}\psi_{4}(r)+k\psi_{5}(r)&=&0,\label{DKP5-1}\\
\,\,\,\,\left(-qA_{t}+\mathcal{E}\right)\psi_{1}(r)-M\psi_{2}(r)&=&0,\label{DKP5-2}\\
\,\,\,\,\,\,\,i\partial_{r}\psi_{1}(r)-M\psi_{3}(r)&=&0,\label{DKP5-3}\\
-\frac{m}{r}\psi_{1}(r)-\frac{i\beta}{r}\partial_{r}\psi_{1}(r)-M\psi_{4}(r)&=&0,\label{DKP5-4}\\
-k\psi_{1}(r)-M\psi_{5}(r)&=&0.\label{DKP5-5}
\end{eqnarray}
\end{subequations}
Next we decouple these equations, we obtain
\begin{subequations}
\begin{eqnarray}
\psi_{2}(r)&=&\frac{1}{M}\left(-qA_{t}+\mathcal{E}\right)\psi_{1}(r),\label{dcoupDKP5-2}\\
\psi_{3}(r)&=&\frac{i}{M}\partial_{r}\psi_{1}(r),\label{dcoupDKP5-3}\\
\psi_{4}(r)&=&-\frac{1}{M}\left(\frac{m}{r}+\frac{i\beta}{r}\partial_{r}\right)\psi_{1}(r),\label{dcoupDKP5-4}\\
\psi_{5}(r)&=&-\frac{k}{M}\psi_{1}(r).\label{dcoupDKP5-5}
\end{eqnarray}
\end{subequations}
From now on, let us consider $qA_{t}=V(r)$. After straightforward algebra by using  Eqs. \eqref{DKPSpinor}, \eqref{dcoupDKP5-2}, \eqref{dcoupDKP5-3}, \eqref{dcoupDKP5-4} and \eqref{dcoupDKP5-5}, we rewrite Eq. \eqref{J0} in the form of
\begin{equation}\label{J00}
J^0=\frac{1}{M}\left(\mathcal{E}-V(r)\right)|\psi_{1}(r)|^2.
\end{equation}
Then, we substitute Eqs. \eqref{dcoupDKP5-2}, \eqref{dcoupDKP5-3}, \eqref{dcoupDKP5-4} and
\eqref{dcoupDKP5-5} into Eq. \eqref{DKP5-1}, and we acquire a radial
differential equation for the first component of the DKP spinor:
\begin{equation}\label{non-coupDKP1}
\begin{split}
&\left(1+\frac{\beta^2}{r^2}\right)\frac{\mathrm{d}^2\psi_{1}(r)}{\mathrm{d}r^2}+\left(\frac{1}{r}-\frac{\beta^2}{r^3}-\frac{2im\beta}{r^2}\right)\frac{\mathrm{d}\psi_{1}(r)}{\mathrm{d}r}\\
&+\left[(V(r)-\mathcal{E})^2-k^2-M^2-\frac{m^2}{r^2}+\frac{im\beta}{\alpha^2 r^3}\right]\psi_{1}(r)=0.
\end{split}
\end{equation}
In order to solve this second-order differential equation, we consider the form of wave functions as \cite{MaiaEPJC2019}
\begin{equation}\label{WF2}
\psi_{1}(r)=e^{im\,tan^{-1}\left(\frac{r}{\beta}\right)}G\left(r\right).
\end{equation}
Here, $G(r)$ is an unknown function and to determine it we utilize Eq. \eqref{WF2} in Eq. \eqref{non-coupDKP1}. We find
\begin{equation}\label{non-coupDKP2}
\begin{split}
&\left(1+\frac{\beta^2}{r^2}\right)\frac{\mathrm{d}^2G(r)}{\mathrm{d}r^2}+\left(\frac{1}{r}-\frac{\beta^2}{r^3}\right)\frac{\mathrm{d}G(r)}{\mathrm{d}r}\\
&+\left[(V(r)-\mathcal{E})^2-k^2-M^2-\frac{m^2}{r^2+\beta^2}\right]G(r)=0.
\end{split}
\end{equation}
Then, we define a new variable as
\begin{equation}\label{var3}
\rho=\sqrt{r^2+\beta^2}.
\end{equation}
If we employ Eq. \eqref{var3} in Eq. \eqref{non-coupDKP2}, we arrive at
\begin{equation}\label{non-coupDKP3}
\frac{\mathrm{d}^2G(\rho)}{\mathrm{d}\rho^2}+\frac{1}{\rho}\frac{\mathrm{d}G(\rho)}{\mathrm{d}\rho}+\left[(V(\rho)-\mathcal{E})^2-k^2-M^2-\frac{m^2}{\rho^2}\right]G(\rho)=0.
\end{equation}
Hereafter, we will investigate analytic solutions for Eq. \eqref{non-coupDKP3}.


\section{Exact solution of the generalized DKP equation\label{sec3}}

In this section, for two different picks of the potential energy, namely in the absence and the presence of a static potential $V(r)$,  we solve the generalized DKP equation in the space-time with a spiral dislocation. To be more concrete, we obtain the wave functions and energy eigenvalues associated with each of these cases by utilizing the Bessel's equation and the NU method.
\begin{enumerate}
\item
{\bf{The first case:}} In this case, we solve Eq. \eqref{non-coupDKP3} in the absence of a static potential $V(r)$. This means that we ignore the minimal coupling embedded in the covariant derivative belonging to the generalized DKP equation. Hence, Eq. \eqref{non-coupDKP3} can be rewritten as follows
\begin{equation}\label{BesselEq}
\frac{\mathrm{d}^2G\left(\rho\right)}{\mathrm{d}\rho^2}+\frac{1}{\rho}\frac{\mathrm{d}G\left(\rho\right)}{\mathrm{d}\rho}+\frac{1}{\rho^2}\left[\varkappa^2\rho^2-\vartheta^2\right]G\left(\rho\right)=0,
\end{equation}
where
\begin{subequations}
\begin{align}
&\varkappa^2\equiv \mathcal{E}_{\vartheta s}^2-k^2-M^2,\\
&\vartheta^2\equiv m^2.
\end{align}
\end{subequations}
It is well-known that Eq. \eqref{BesselEq}  is related to the Bessel differential equation so that the general solution of Eq. \eqref{BesselEq} is given by
\begin{equation}\label{GBS}
G\left(\rho\right)=A J_{|\vartheta|}\left(\varkappa\rho\right)+B N_{|\vartheta|}\left(\varkappa\rho\right),
\end{equation}
in which $A$ and $B$ are constants, and also $J_{|\vartheta|}\left(\varkappa\rho\right)$ is the the Bessel functions of the first kind and $N_{|\vartheta|}\left(\varkappa\rho\right)$ is the Bessel functions of the second kind. Because of the diverging of the Bessel function of the second kind at the origin, we need to choose $B=0$ in Eq. \eqref{GBS}, thus, we have a regular solution at the origin \cite{MaiaEPJC2019,ZareGRG2020}. Then, we express the wave function as follows
\begin{equation}\label{WFBF}
G_{m}\left(r\right)=AJ_{|m|}\left(\sqrt{\left(\mathcal{E}_{\vartheta s}^2-k^2-M^2\right)\left(r^2+\beta^2\right)}\right),
\end{equation}
where $A$ is the normalization constant. According to Eqs. \eqref{J00} and \eqref{WFBF}, we express the charge density in the form of
\begin{equation}\label{J0BF}
J^0=\frac{A^2}{M}\left(\mathcal{E}_{\vartheta s}-V(r)\right)\left|J_{|m|}\left(\sqrt{\left(\mathcal{E}_{\vartheta s}^2-k^2-M^2\right)\left(r^2+\beta^2\right)}\right)\right|^2.
\end{equation}
To find energy eigenvalues, we need to vanish the wave function, so that,  by considering the hard-wall confining condition, the wave function vanishes at $r_0$, that is, $J_{|\vartheta|}\left(\varkappa\rho_{0}\right)=0$, this means that the wave function $J_{|\vartheta|}\left(\varkappa\rho\right)$ has to fade away when $\rho\rightarrow\rho_{0}=\sqrt{r_{0}^2+\beta^2}$. Therefore, in order to get energy quantization, we have to consider $\varkappa \sqrt{r_{0}^2+\beta^2}=x_{\vartheta,s}$, in which $x_{\vartheta,s}$ is the $s$th zero of the Bessel function $J_{|\vartheta|}\left(\varkappa\rho\right)$. After all, we obtain the energy eigenvalue function in the form of
\begin{equation}\label{energyBF}
\mathcal{E}_{\vartheta s}=\pm\sqrt{\frac{x_{\vartheta,s}^2}{r_{0}^2+\beta^2}+k^2+M^2}
\end{equation}
As we notice the energy eigenvalues have both signs. Furthermore, the quantity inside of the square root is definitely positive. Thereby, we conclude that this bosonian system allows positive and negative solutions of the energy eigenvalues. Furthermore, we observe out of Eq. \eqref{energyBF} that the energy eigenvalues do not depend on the angular momentum quantum number $m$.

Next we use numerical values to represent our findings graphically. During the whole calculations in this first case, we assume $k=1$, $M=1$, $x_{0,1}=2.4048$ and $m=0$. First, we intend to observe the effect of the defect parameter $\beta$ on the charge density. Therefore, we illustrate the charge density versus the distance for three different values of deformation parameter, namely $\beta=0.1, 0.5$, and $0.9$ in Fig. \ref{fig1}. We take
$r_{0}=0.5$ to calculate the normalization constants. For each deformation parameter values we obtained the normalization constants and presented their values.  Then, by using  Eq. \eqref{J0BF} we plot $J^0$ via $r$. In Fig. \ref{fig11}, the highest peak is the one which is close to the origin. It is obtained for the smallest deformation value. As seen from Fig. \ref{fig12}, the highest probability of detecting the spin-zero particle is around the origin. We observe the deformation parameter values change the probabilities of observing the DKP particles near the origin. In Fig. \ref{fig13} we realize that the confinement probability are mainly divided into two regions.  When $\beta=0.9$ the confinement probability around the origin nearly vanishes.
\begin{figure}[H]
	\centering
	\subfigure[With three different values of $\beta$]{%
		\label{fig11}%
		\includegraphics[height=6cm,width=7cm]{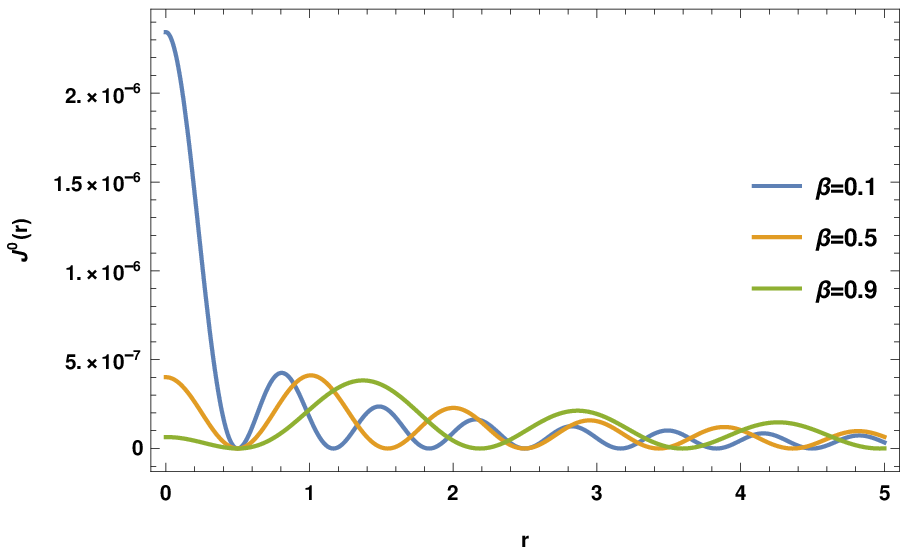}}%
	\qquad
	\subfigure[$\beta=0.1$ and $A=73\times 10^{-5}$]{%
		\label{fig12}%
		\includegraphics[height=6cm]{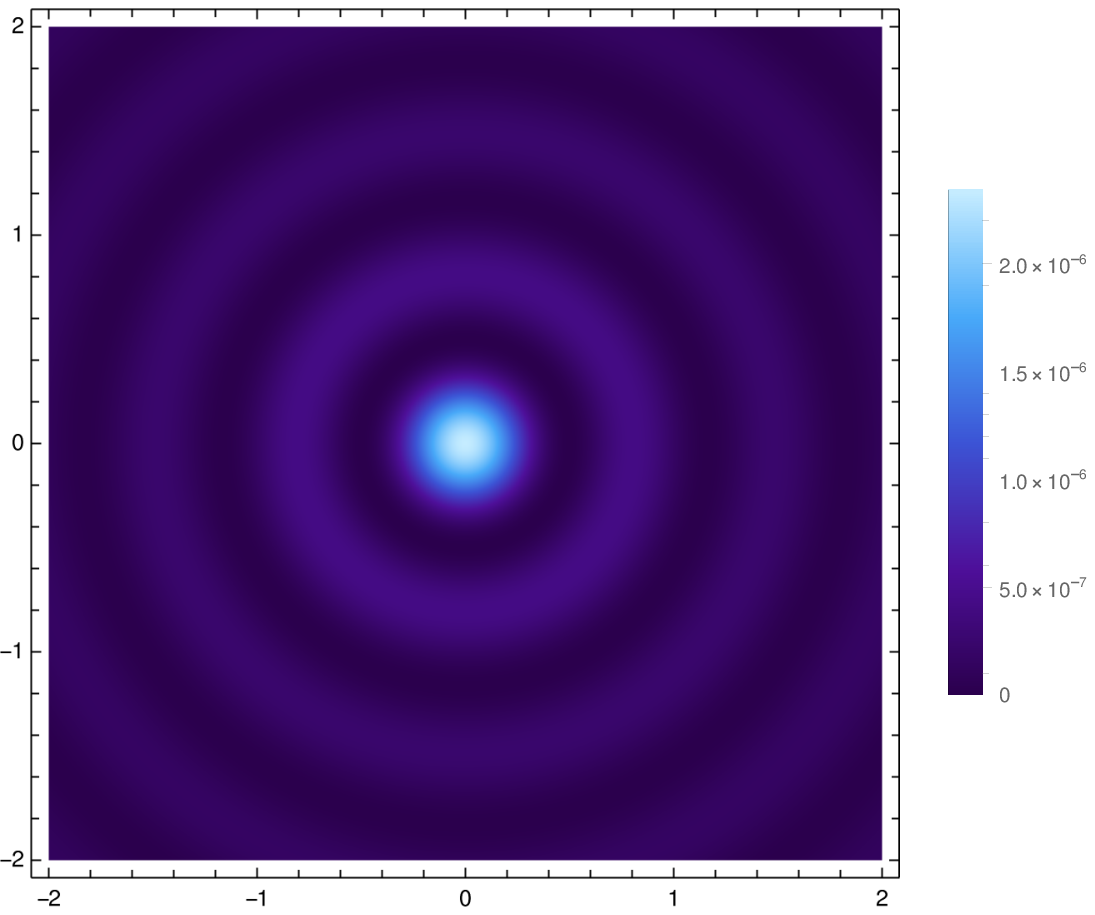}}%
		
	\subfigure[$\beta=0.5$ and $A=83\times 10^{-5}$]{%
		\label{fig13}%
		\includegraphics[height=6cm]{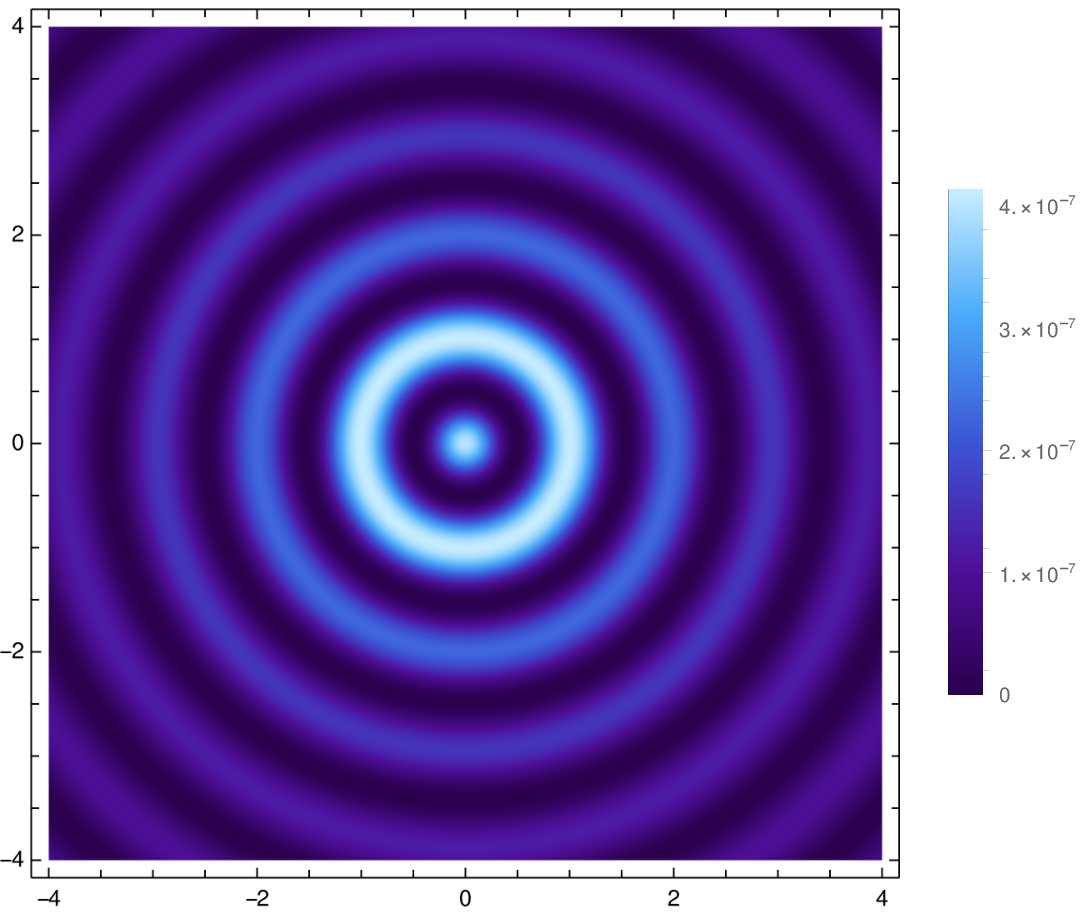}}%
	\qquad
	\subfigure[$\beta=0.9$ and $A=93\times 10^{-5}$]{%
		\label{fig14}%
		\includegraphics[height=6cm]{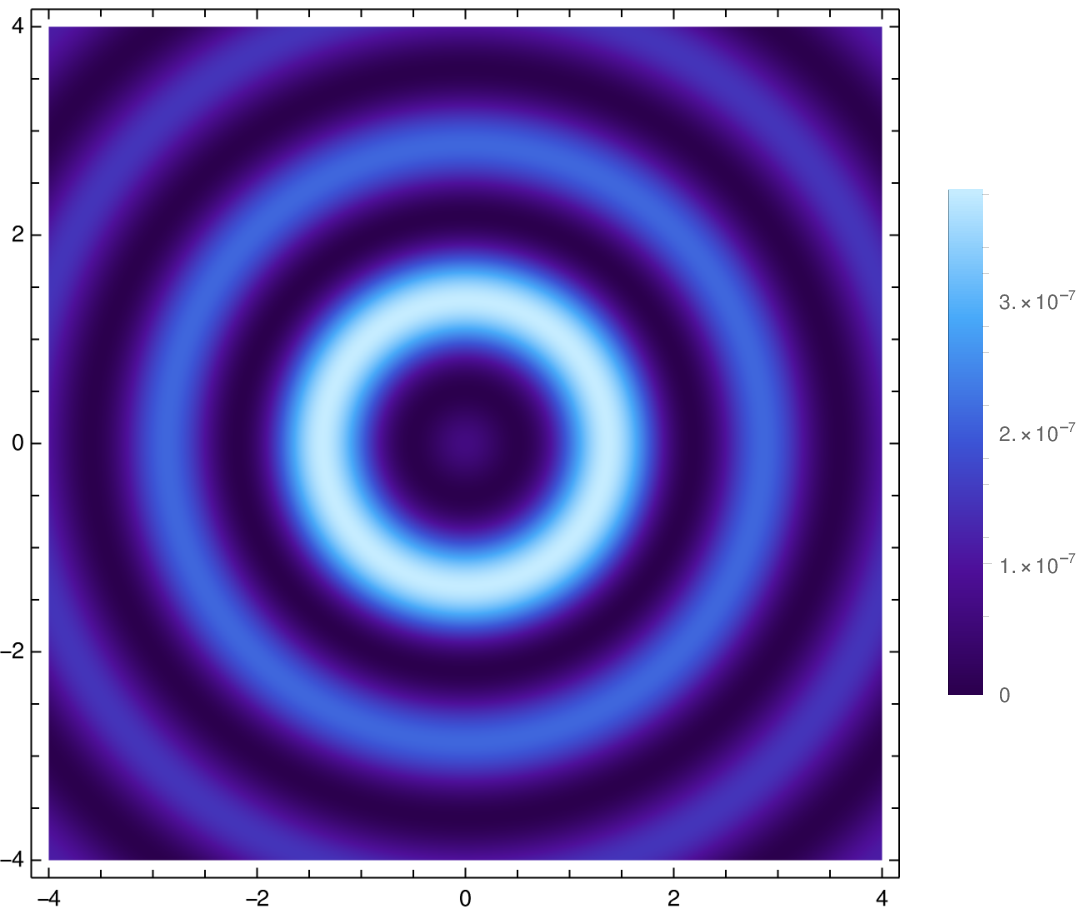}}%
\caption{The wave function
probability density versus the distance in the defected
space-time.}\label{fig1}
\end{figure}
Then, we investigate the effect of the $\beta$ parameter on the energy eigenvalues.
Here, we assign three different values to deformation parameter and depict $\mathcal{E}_{\vartheta s}$ versus $r_0$ in Fig. \ref{fig2}. Around $r_0=0$, we find
$\mathcal{E}_{01}(\beta=0.6)<\mathcal{E}_{01}(\beta=0.3)<\mathcal{E}_{01}(\beta=0.1)$. We observe a decrease in the energy eigenvalues while we increase $r_0$ from zero to one. We find out that this decrease is faster for small parameters. We see that the energy eigenvalues tend to the same value, therefore we conclude that they become degenerate around $r_0=1$.
\begin{figure}[H]
	\centering
	\subfigure[With three different values of $\beta$]{%
		\label{fig21}%
		\includegraphics[height=6cm,width=7cm]{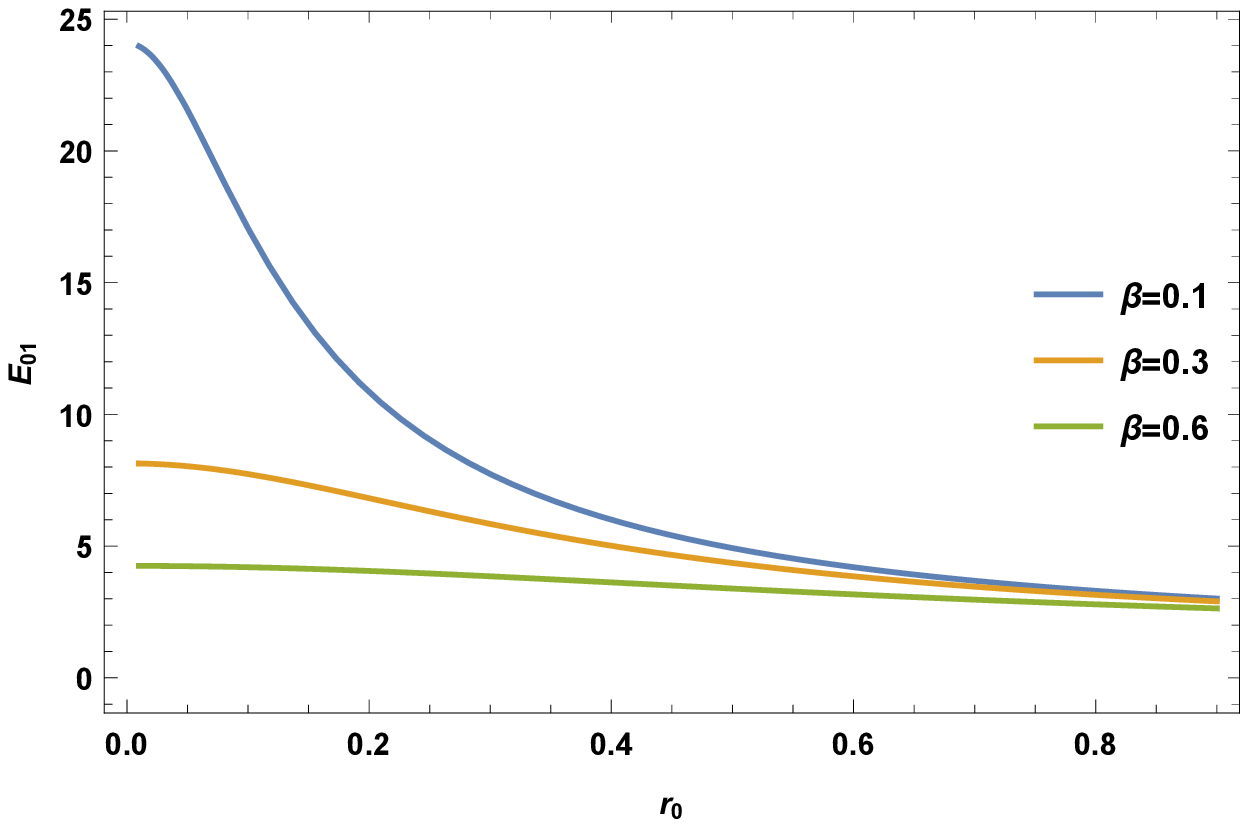}}%
	\qquad
	\subfigure[$\beta=0.1$]{%
		\label{fig22}%
		\includegraphics[height=6cm]{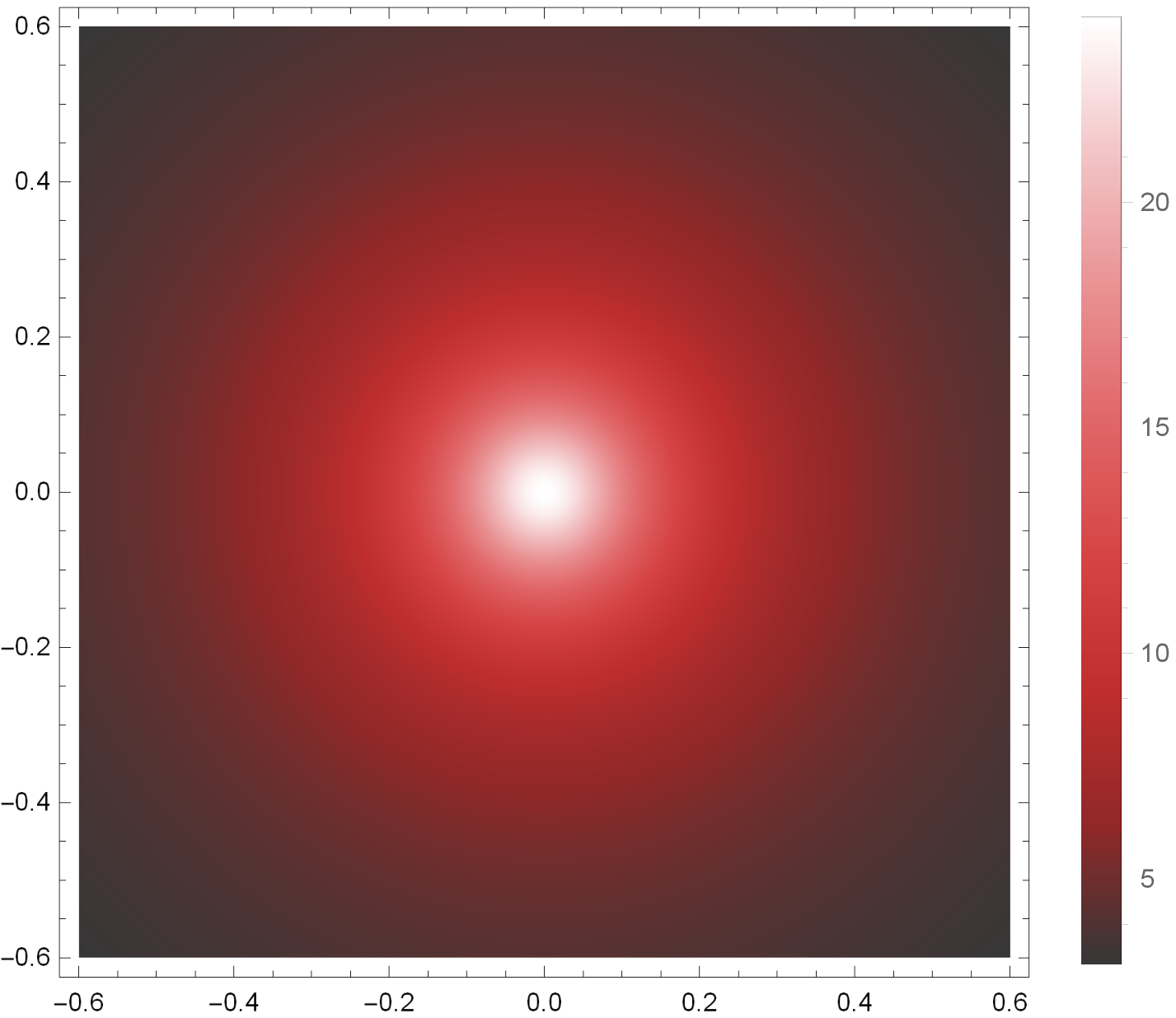}}%
	
	\subfigure[$\beta=0.3$]{%
		\label{fig23}%
		\includegraphics[height=6cm]{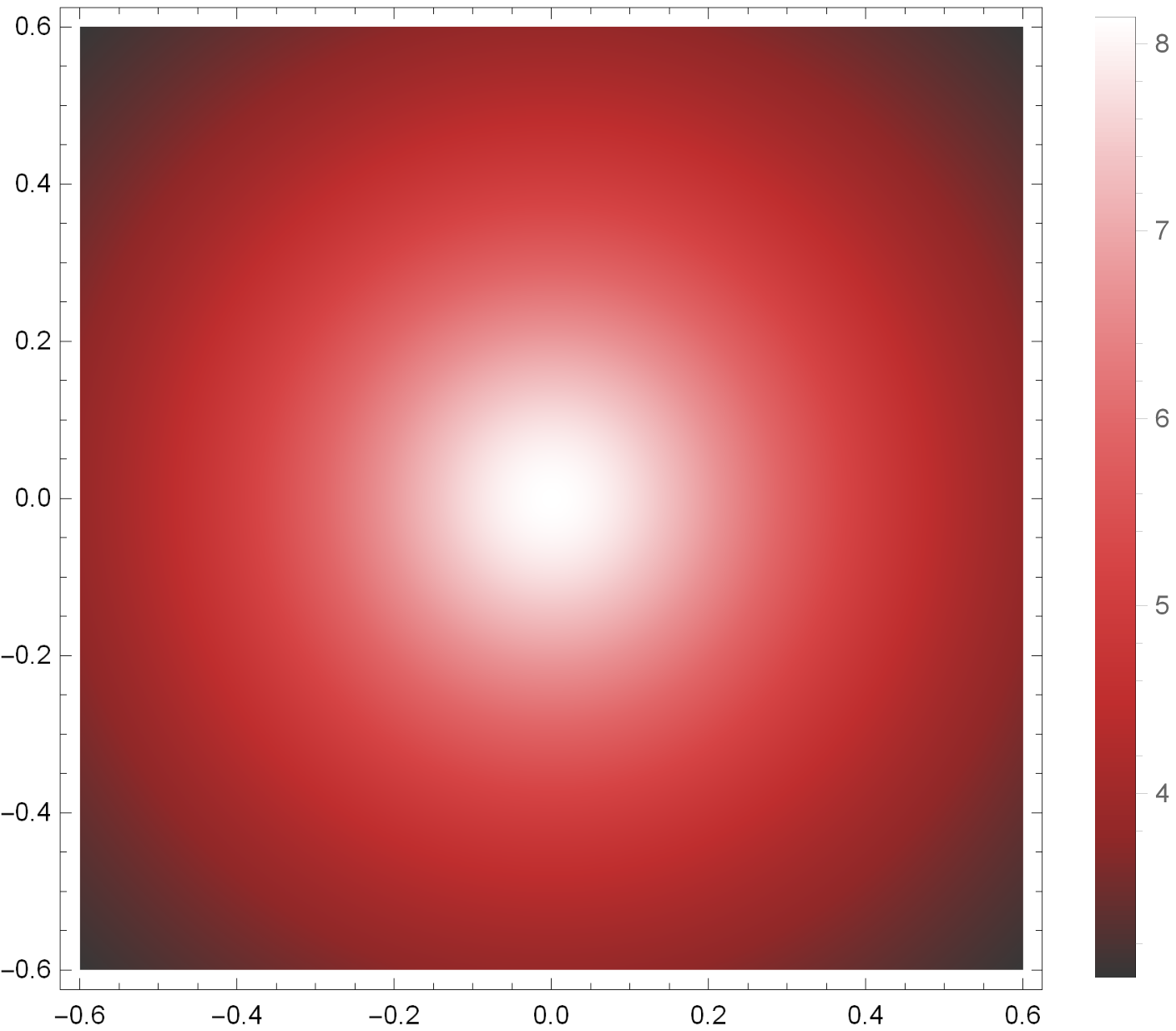}}%
	\qquad
	\subfigure[$\beta=0.6$]{%
		\label{fig24}%
		\includegraphics[height=6cm]{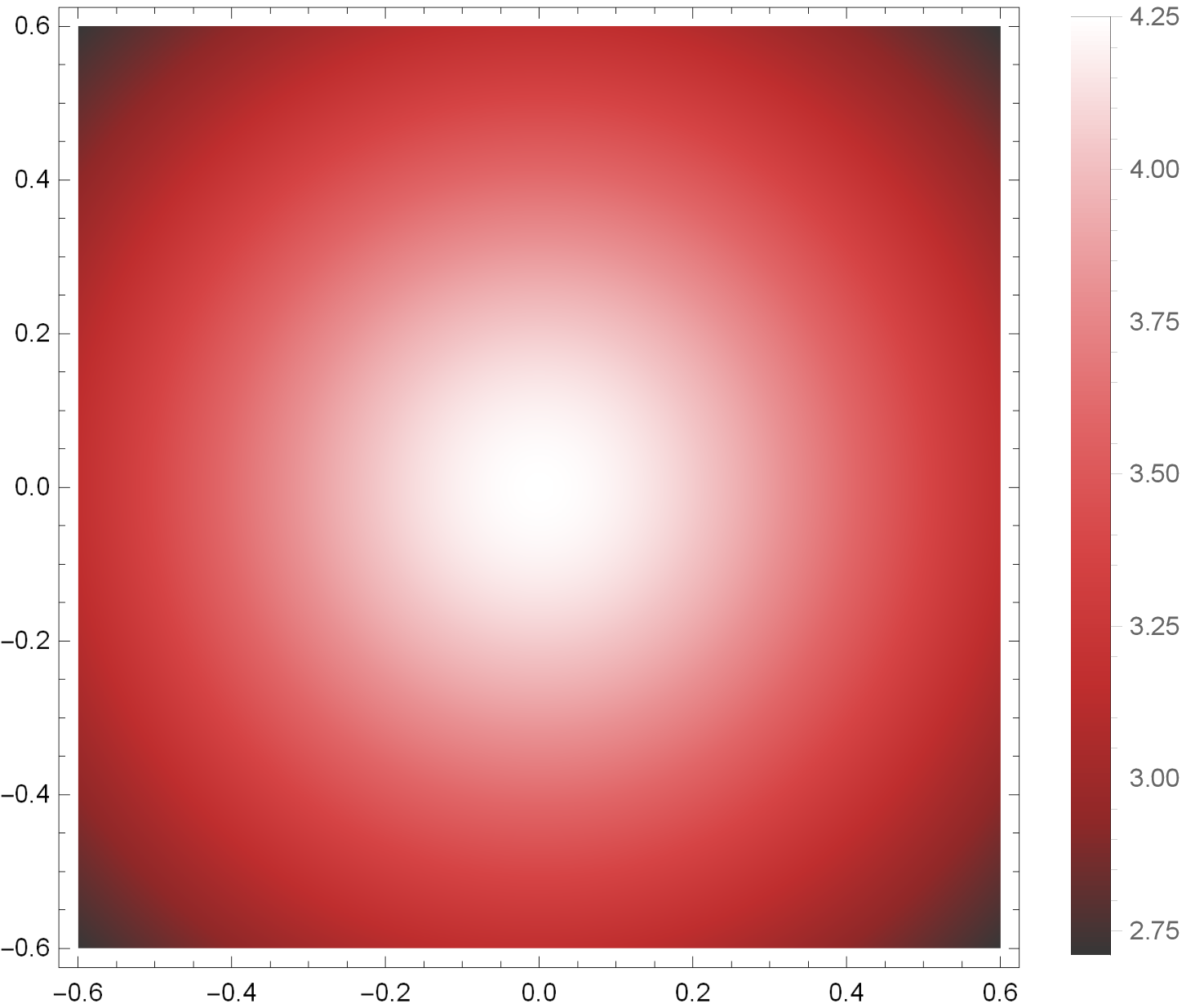}}%
	\caption{The energy eigenvalues, form Eq. \eqref{energyBF}, is plotted as a function of $r_0$ in the defected space-time, according to three different values of $\beta=0.1, 0.3 \,\mathrm{and}\, 0.6$.}
	\label{fig2}
\end{figure}

\item
{\bf{The second case:}}
In this case, we carry out the spin-zero DKP equation in the presence of the Coulomb-like potential so that our choice of potential is
\begin{equation}\label{coulomblikePot1}
V(r)=\frac{\alpha_{c}}{\sqrt{r^2+\beta^2}}.
\end{equation}
Here, the coefficient $\alpha_{c}$ is a constant, and $\beta$ denotes the parameter that is related to the topological defect in space-time. In this potential, If one takes $\beta=0$, the Coulomb-like potential reduces to the well-known Coulomb potential, while for $\beta\ne0$  the Coulomb-like potential turns to soft-core Coulomb potential \cite{Hall2009,Hall2010}. In the present case, according to the variable given in Eq. \eqref{var3}, the  Coulomb-like potential can be considered as
\begin{equation}\label{coulomblikePot2}
V(\rho)=\frac{\alpha_{c}}{\rho}.
\end{equation}
We continue by replacing the potential Eq. \eqref{coulomblikePot2} in Eq. \eqref{non-coupDKP3}. We arrive at
\begin{equation}\label{NUEq}
\frac{\mathrm{d}^2G\left(\rho\right)}{\mathrm{d}\rho^2}+\frac{1}{\rho}\frac{\mathrm{d}G\left(\rho\right)}{\mathrm{d}\rho}+\frac{1}{\rho^2}\left[-\left(k^2+M^2-\mathcal{E}_{nm}^2\right)\rho^2-2\alpha_{c}\mathcal{E}_{nm}\rho+\alpha_{c}^2-m^2\right]G\left(\rho\right)=0,
\end{equation}
which is known as the NU differential equation. Thus, according to Refs. \cite{TS,deMontigny},  the wave function of Eq. \eqref{NUEq} can be presented in terms of the generalized Laguerre polynomial  as follows
\begin{equation}\label{WFCL}
G_{nm}\left(r\right)=\mathrm{N}_{nm}\left(r^2+\beta^2\right)^{\alpha_{12}/2}e^{\alpha_{13}\sqrt{r^2+\beta^2}}L_{n}^{\alpha_{10}-1}\left(\alpha_{11}\sqrt{r^2+\beta^2}\right).
\end{equation}
Here, $\mathrm{N}_{nm}$ indicates the normalization constant. Then, we obtain the parameters $\alpha_{10}$, $\alpha_{11}$, $\alpha_{12}$ and $\alpha_{13}$  as
\begin{equation}
\begin{split}
&\alpha_{10}=1+2\sqrt{m^2-\alpha_{c}^2},\qquad \alpha_{11}=2\sqrt{M^2+k^2-\mathcal{E}_{nm}^2},\\
&\alpha_{12}=\sqrt{m^2-\alpha_{c}^2},\qquad\quad\,\,\,\, \alpha_{13}=-\sqrt{M^2+k^2-\mathcal{E}_{nm}^2}.
\end{split}
\end{equation}
With respect to Eqs. \eqref{J00} and \eqref{WFCL}, the charge density can be rewritten as
\begin{equation}\label{J0CL}
J^0=\frac{\mathrm{N}_{nm}^2}{M}\left(\mathcal{E}_{nm}-V(r)\right)\left|\left(r^2+\beta^2\right)^{\alpha_{12}/2}e^{\alpha_{13}\sqrt{r^2+\beta^2}}L_{n}^{\alpha_{10}-1}\left(\alpha_{11}\sqrt{r^2+\beta^2}\right)\right|^2,
\end{equation}
 while the energy eigenvalues $\mathcal{E}_{nm}$ are derived as
\begin{equation}\label{energyNU}
\mathcal{E}_{nm}=\pm\left|1+2n+2\sqrt{m^2-\alpha_{c}^2}\right|\sqrt{\frac{M^2+k^2}{4\alpha_{c}^2+\left(1+2n+2\sqrt{m^2-\alpha_{c}^2}\right)^2}}.
\end{equation}
Alike the previous case, this bosonian system also allows positive and negative solutions for the energy eigenvalues. As a difference, Eq. \eqref{energyNU} shows that the energy eigenvalues depend on the  quantum numbers $n$ and $m$.

In Fig. \ref{fig3}, we depict the probability density as a function distance. We take $k=1, M=1, \beta=0.5, \alpha_c=-0.5$ and $m=1$ and calculate the normalization constants for three different quantum numbers. We observe that when the quantum numbers increase, the probability of detecting the DKP particle near the origin decreases as shown in Fig. \ref{fig31}. Then, we illustrate the confinement probabilities in Fig. \ref{fig32} for $n=1$, in Fig. \ref{fig33} for $n=3$, and in Fig. \ref{fig34} for $n=6$,
\begin{figure}[H]
	\centering
	\subfigure[Whit three different values of $n$]{%
		\label{fig31}%
		\includegraphics[height=6cm,width=7cm]{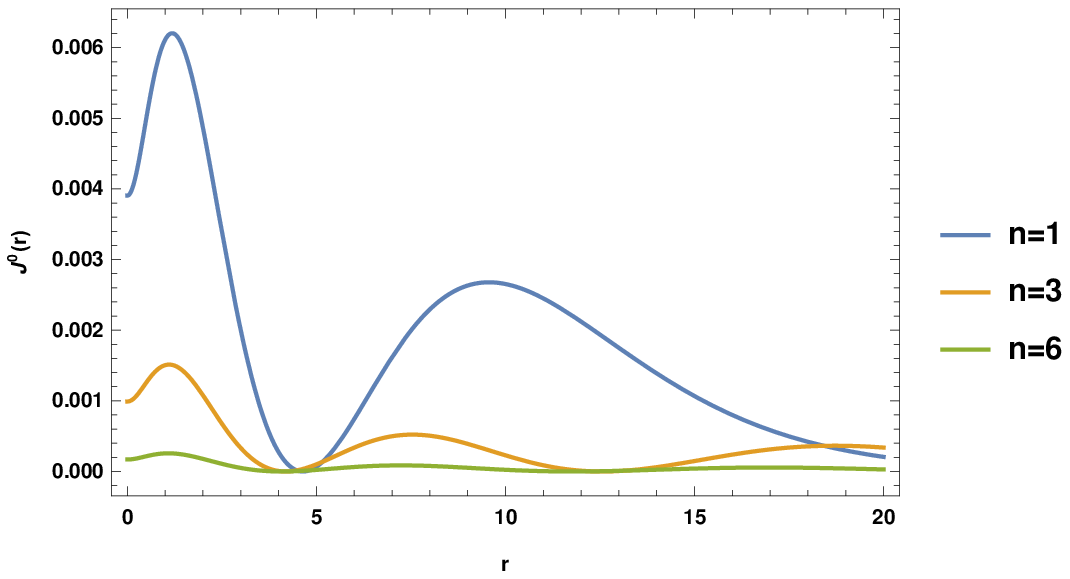}}%
	\qquad
	\subfigure[$n=1$ and $\mathrm{N}_{11}=35\times10^{-3}$]{%
		\label{fig32}%
		\includegraphics[height=6cm]{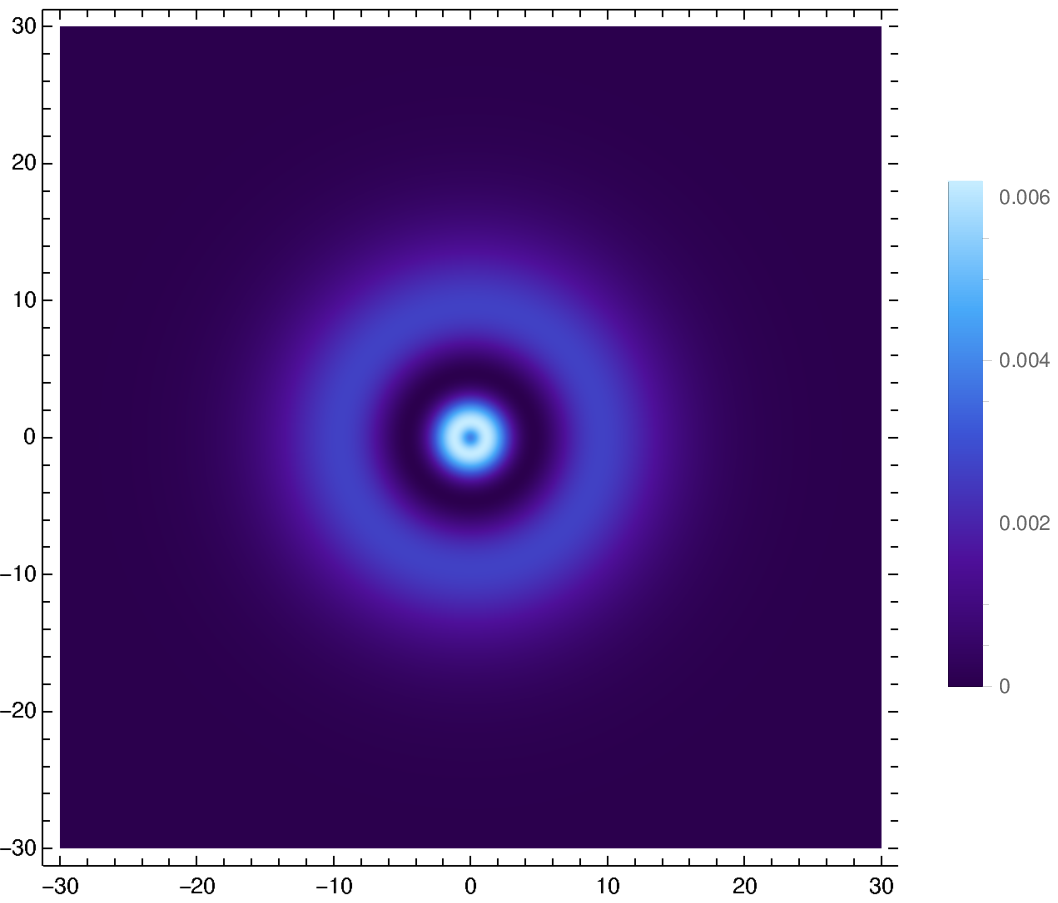}}%
	
	\subfigure[$n=3$ and $\mathrm{N}_{31}=6\times10^{-3}$]{%
		\label{fig33}%
		\includegraphics[height=6cm]{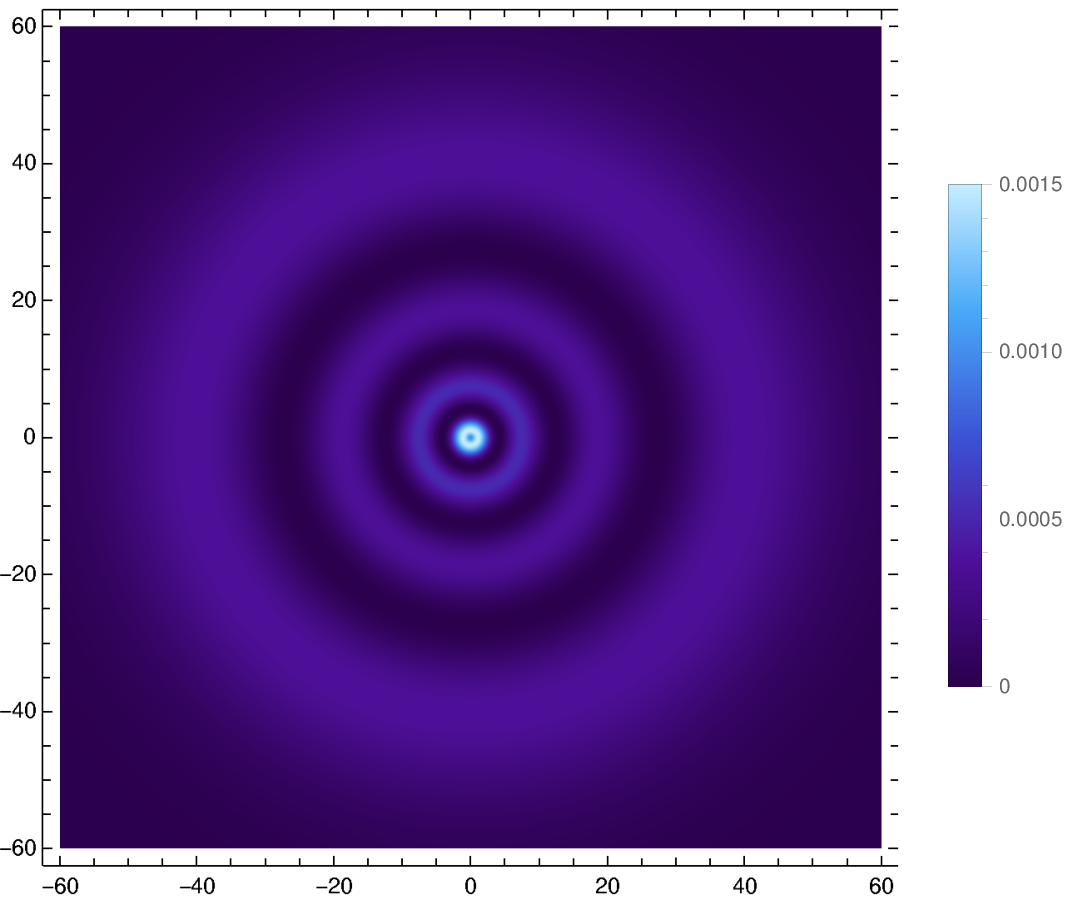}}%
	\qquad
	\subfigure[$n=6$ and $\mathrm{N}_{61}=1\times10^{-3}$]{%
		\label{fig34}%
		\includegraphics[height=6cm]{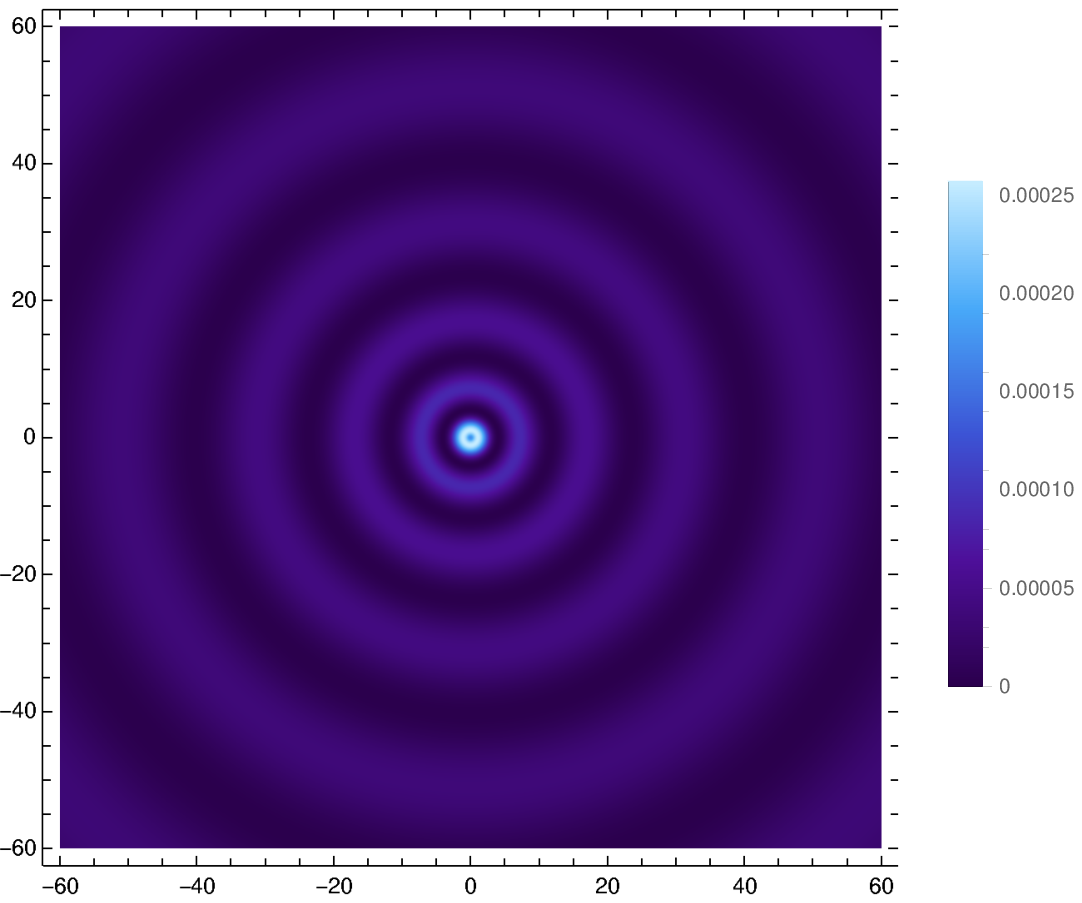}}%
	\caption{The wave function
		probability density, form Eq. \eqref{WFCL}, is plotted as a function of $r$ in the defected spacetime.}\label{fig3}
\end{figure}
Finally, we present some comments on the energy eigenvalue function that is given in Eq. \eqref{energyNU}. It is worth noting that the energy eigenvalues, unlike the wave function  do not depend on the $\beta$ parameter. Then, we plot energy eigenvalue function versus the potential parameter, $\alpha_{c}$, in Fig. \ref{fig41}. Moreover, we take  $k=1$, $M=1$, $\beta=0.5$ and illustrate the dependence of the energy eigenvalue on the quantum number $m$ in Fig. \ref{fig42}. In both graphs, we obtain a similar behaviour, namely with the increase of the parameters, the energy eigenvalues tend to one common value.
\begin{figure}[H]
	\centering
	\subfigure[For $m=1$]{%
		\label{fig41}%
		\includegraphics[height=6cm,width=7cm]{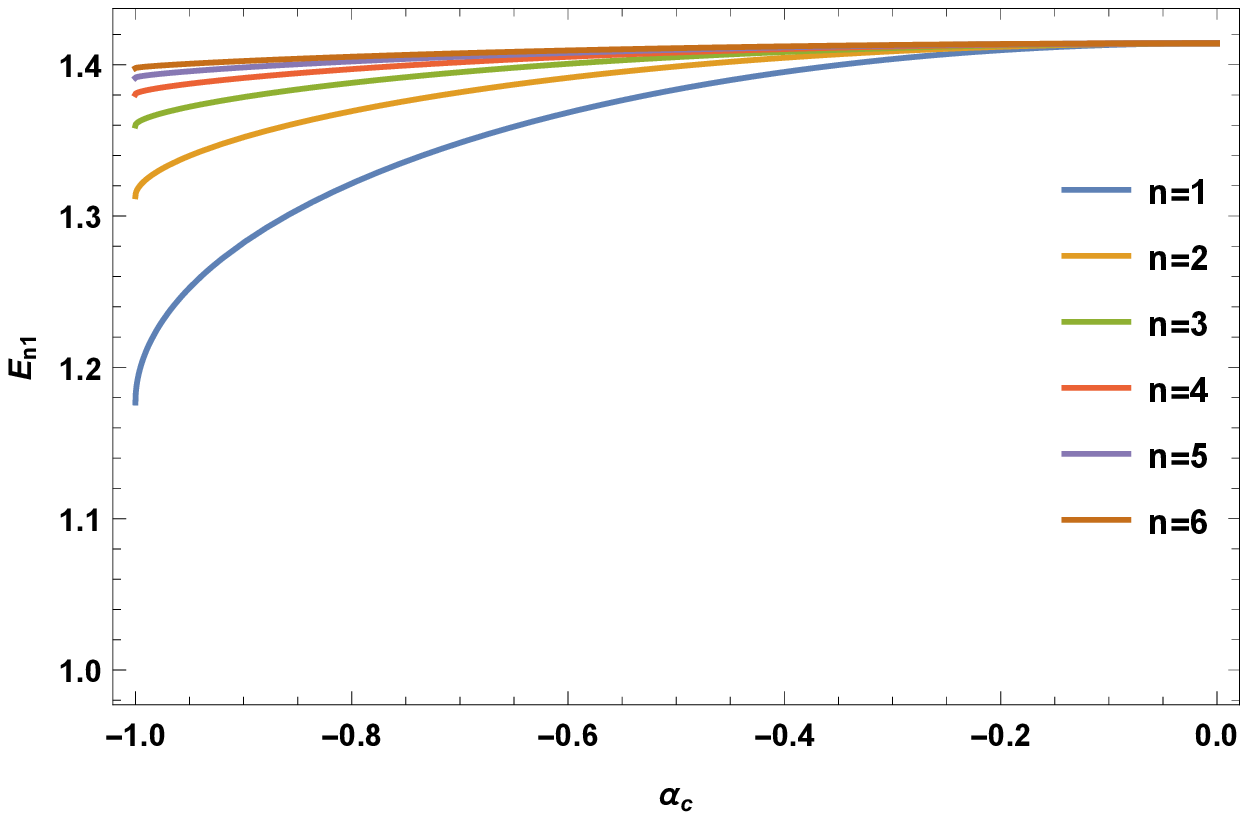}}%
	\qquad
	\subfigure[For $\alpha_{c}=-0.2$]{%
		\label{fig42}%
		\includegraphics[height=6cm,width=7cm]{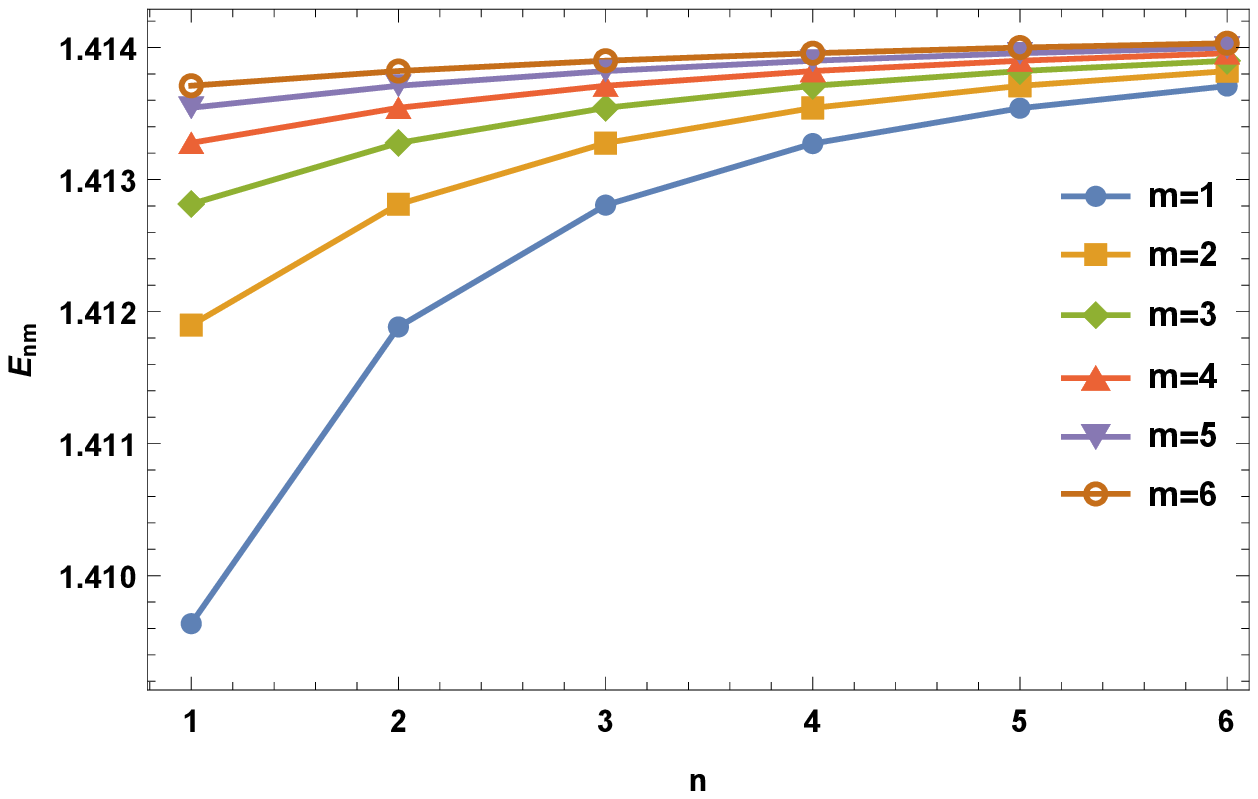}}%
	\caption{The energy eigenvalues, form Eq. \eqref{energyNU}, is plotted in terms of $\alpha_{c}$ and $n$ in the defected space-time, according to six different values of $n$ and $m$.}
	\label{fig4}
\end{figure}
\end{enumerate}



\section{Conclusion\label{Conc}}
In this study, we present the introduction of a line element of the spiral dislocation, and then,  we derive the generalized spin-zero DKP equation under the background of the considered spiral dislocation. After all, we obtain a second-order differential equation by using appropriate changing variables. Then, we investigate solutions for the generalized DKP equation in the absence and in the presence of static potential $V(r)$, respectively. In both cases, we find energy eigenvalues and corresponding wave functions with analytical methods. In the first case where the scalar potential energy does not exist, we observe that the wave function depends on the quantum number $m$ and the parameter $\beta$ which is associated with the defect. However, the energy eigenvalue function depends only on the deformation parameter. When we consider a Coulomb-like scalar potential energy, as the second case, we observe that the energy eigenvalue function and the corresponding wave function depend on the quantum numbers $n$ and $m$. In this case, the deformation parameter is merely seen in the wave function. By assigning random numerical values to the parameters, we strengthened our findings by presenting them graphically.

\section*{Acknowledgment}
The authors thank the referee for a thorough reading of our manuscript and for constructive suggestion. This work is supported by the Internal Project, $[2020/2209]$, of Excellent Research of the Faculty of Science of University Hradec Kr\'{a}lov\'{e}. One of the author, B.C. L\" utf\"uo\u{g}lu, was partially supported by the Turkish Science and Research Council (T\"{U}B\.{I}TAK).









\end{document}